\documentclass{elsart1p}

\usepackage{amsmath}
\input{epsf}

\newcommand{\pimass}{m_{\pi}}

\newcommand{\pimassrhosq}{m_{\pi}^{*2}}
\newcommand{\Nmass}{m_{N}}

\renewcommand{\cal}{\mathcal}
\usepackage[usenames]{color}
\usepackage{ulem} 

\renewcommand{\em}{\it}

\begin{document}

\begin{frontmatter}
\begin{flushright}
YITP-07-23,\,  KUNS-2142
\end{flushright}


\title{ 
In-medium Pion and Partial Restoration of Chiral Symmetry}

\author[YITP]{D. Jido},
\author[TU]{T. Hatsuda},
\author[YITP,PD]{T. Kunihiro}

\address[YITP]{Yukawa Institute for Theoretical Physics, Kyoto University, Kyoto, 606-8502, Japan}
\address[TU]{Department of Physics, The University of Tokyo, Tokyo, 113-0033, Japan
}
\address[PD]{Department of Physics, Kyoto University, Kyoto, 606-8502,
Japan}

\date{\today}

\begin{abstract}
The partial restoration of chiral symmetry in nuclear medium is 
investigated in a model independent way by exploiting operator relations
in QCD. 
An exact sum rule is derived for the quark condensate valid for all density. 
This sum rule is simplified at low density to 
a new relation
{$\langle \bar{q}q \rangle^*/\langle \bar{q}q \rangle =
(F_{\pi}^t/F_{\pi}) Z_{\pi}^{*1/2}$}
with the in-medium quark condensate $\langle \bar{q}q \rangle^*$,
in-medium pion decay constant $F_{\pi}^t$ 
and in-medium pion wave-function renormalization 
$Z_{\pi}^{*}$.
Calculating $Z_{\pi}^{*}$
at low density 
from the iso-scalar pion-nucleon scattering data and 
relating $F_{\pi}^t$ to the isovector pion-nucleus scattering length
$b_1^*$,  it is concluded that 
the enhanced repulsion of the s-wave isovector pion-nucleus interaction observed 
in the deeply bound pionic atoms directly implies the
reduction of the in-medium quark condensate. The
knowledge of the in-medium pion mass $m_{\pi}^*$ 
is not necessary to reach this conclusion.
\end{abstract}

\end{frontmatter}

\section{Introduction}

Exploring possible evidence of partial restoration of chiral symmetry in 
the nuclear medium is one of the 
most important and challenging problems in hadron physics.  
Experimental studies along this line have been carried out 
by the spectroscopy of deeply bound pionic atoms \cite{Suzuki:2002ae},
by low energy pion-nucleus scatterings  \cite{Friedman:2004jh}, 
and by the production of di-pions  in hadron-nucleus and photon-nucleus
reactions \cite{Bonutti:2000bv,Starostin:2000cb,Messchendorp:2002au}.
Important experimental observations are 
(i)  a {\em repulsive enhancement} of the in-medium 
$\pi^{-}$-nucleon interaction in the pionic atoms and pion-nucleus scattering, and  
(ii) an {\em attractive enhancement} of the scalar-isoscalar $\pi$-$\pi$ interaction
in nuclei.
In the theoretical side,
it was suggested that the reduction of the 
temporal part of the pion decay constant in the nuclear medium
$F_{\pi}^t$ is intimately related to the phenomenon
(i) \cite{Kolomeitsev:2002gc,Weise:2005ss}. 
It was also argued that the reduction of $F_{\pi}^t$ is responsible for 
the phenomenon (ii) \cite{Jido:2000bw}.

The purpose of this paper is 
to present theoretical basis of the direct 
connection between $F_{\pi}^t$ and 
the in-medium quark condensate $\langle \bar{q}q \rangle^*$ 
(the order parameter of the chiral phase transition)
only by using the current commutation relations, spectral decomposition
and the linear density approximation.\footnote{A preliminary account of this work 
has been reported in  \cite{Jido:2007yt}.}
We derive an exact sum rule  relating the in-medium 
quark condensate and hadronic matrix elements 
valid for all density in the chiral limit.
We also derive following model-independent relations 
at low density:
The im-medium Glashow-Weinberg (GW) relation  
$F_{\pi}^t G_{\pi}^{*1/2} = - \langle \bar{q}q \rangle^*$ in the chiral
limit,
the in-medium Tomozawa$-$Weinberg (TW) relation 
${\cal T}^{(-)*}(\omega) \simeq \omega/(2(F_{\pi}^{t})^{2})$
in the chiral limit,
and the in-medium Gell-Mann$-$Oakes$-$Renner (GOR) relation 
$ ( F^{t}_{\pi})^{2} \pimassrhosq   = - 2 m_{q} \langle \bar q q \rangle^{*}$
away from the chiral limit.
Here
$G_{\pi}^*$ is the pseudo-scalar pion coupling constant,
${\cal T}^{(-)*}(\omega)$ is the isovector pion-nucleus scattering amplitude 
at zero spacial momentum
and $m_{\pi}^*$ is the isospin averaged pion mass. 
In particular, the in-medium GW relation, which
has not been discussed in previous literatures, 
gives a new way to estimate $\langle \bar{q}q \rangle^*$ from 
the experimental data of the pionic atoms and the
pion-nucleus and pion-nucleon
scatterings without recourse to the information of $m_{\pi}^*$ as
will be shown in this paper.

The organization of this paper is as follows:
In section \ref{sec:exact}, we derive the exact sum rule for the 
quark condensate valid for all density in the chiral limit.
In section \ref{sec:GW}, we show that the exact sum rule
leads to the in-medium GW relation and the 
in-medium GOR relation which are valid at low density.
In section \ref{sec:WF}, we evaluate the reduction of $G_{\pi}^*$ 
by relating it to the isoscalar $\pi$N scattering amplitude.
In section \ref{sec:TW}, we derive the in-medium TW relation in the 
chiral limit.  In section \ref{sec:QC},  utilizing all the results derived,
the in-medium quark condensate is related to the observables in 
$\pi$-nucleus and $\pi$N scattering data. 
{In section \ref{sec:offCL}, we discuss how the exact sum rule 
is expressed off the chiral limit, based on the PCAC relation. 
In section \ref{sec:ChPT} the relation of the present argument to the in-medium
chiral perturbation theory is discussed.}
Section \ref{sec:summary} is devoted to summary and concluding remarks.

\section{Exact sum rule in the chiral limit at all density}
\label{sec:exact}

Let us first discuss how the in-medium quark condensate 
is expressed in terms of hadronic quantities. 
We consider the following correlation function
in isospin symmetric nuclear matter in  the chiral limit:
\begin{equation}
\Pi_5^{ab}(q) 
= \int d^4 x \ e^{iq\cdot x} \partial^\mu \langle \Omega | {\rm T}
[A_\mu^a(x) \phi_5^b(0)] | \Omega \rangle ,
\label{eq:Aphicorr}
\end{equation}
where $\phi_{5}$ is the pseudoscalar density 
$\phi_{5}^{a} \equiv \bar \psi i\gamma_{5} (\tau^{a}/2) \psi$ 
with the Pauli matrix $\tau^{a}$ and the quark field $\psi=(u,d)^{\rm T}$; 
$A_{\mu}^{a}$ denotes the axial vector current associated with the SU(2)
chiral transformation.
The ground state of the isospin symmetric 
nuclear matter denoted by  $|\Omega\rangle$ in Eq.\eqref{eq:Aphicorr} 
is normalized as $\langle \Omega | \Omega \rangle =1$ 
and is specified by the baryon density $\rho$.

Invoking an operator relation 
$
\partial^{\mu} {\rm T} [A_{\mu}^{a}(x) \phi_{5}^{b}(0)]
= {\rm T} [\partial^{\mu} A_{\mu}^{a}(x) \phi_{5}^{b}(0)] 
+ \delta(x_{0}) [A_{0}^{a}(x), \phi_{5}^{b}(0)]
$
with the conservation of the axial current $\partial^{\mu} A^{a}_{\mu}=0$ in the chiral limit, 
$\Pi_5^{ab}(0)$ is 
written in terms of the in-medium 
quark condensate $\langle \bar q q \rangle^{*}$ as
\begin{equation}
\Pi_5^{ab}(0) =
 \langle \Omega |  [Q_{5}^{a}, \phi_{5}^{b}] | \Omega \rangle  
= - i \delta^{ab} \langle \bar q q \rangle^{*}
\ , \label{eq:FD2nd}
\end{equation}
where  $Q_{5}^{a}(t) = \int d^{3}x A_{0}^{a}(t,\vec x\,)$ 
and $\langle \bar q q \rangle^{*} \equiv \langle \Omega | \phi | \Omega \rangle $
with the scalar density $\phi$ defined by
$\phi\equiv \frac{1}{2} \bar \psi  \psi = \frac{1}{2}(\bar{u}u+\bar{d}d)$.

The hadronic contributions to 
$\Pi_5^{ab}(q)$
can be
read off from the  spectral representation:
\begin{equation}
\label{eq:disp}
\! \! \Pi_5^{ab}(q) = -i q^{\mu} \int_{0}^{\infty} d \omega' \left( 
\frac{\sigma_{\mu}^{+}(\omega', \vec q\,)}{\omega-\omega'+i\epsilon} -
\frac{\sigma_{\mu}^{-}(\omega', \vec q\,)}{\omega+\omega'-i\epsilon} \right)\ ,
\end{equation}
where $q=(\omega,\vec q)$ with the energy
$\omega$ measured from the nuclear matter ground state.
The spectral functions $\sigma_{\mu}^{\pm}(\omega, \vec q)$ 
give the strength of the hadronic
excitations with the energy $\omega$ and momentum $\vec q$
and are expressed in terms of 
the matrix elements of $A_{\mu}^{a}$ and $\phi_{5}^{a}$: 
\begin{eqnarray}
\sigma_{\mu}^{+}(\omega,\vec q\,) &=& i\sum_{\ell} \langle \Omega | A_{\mu}^{a}|
\Omega_{\ell} \rangle \langle \Omega_{\ell} | \phi_{5}^{b}| \Omega \rangle \, 
\frac{\delta_{ {\vec q},{\vec p}_{\ell} }
\delta( \varepsilon_{\ell}-\omega)}{2\varepsilon_{\ell}} \ , \\
\sigma_{\mu}^{-}(\omega,\vec q\,) &=& i\sum_{\ell} \langle \Omega |\phi_{5}^{b} |
\Omega_{\ell} \rangle \langle \Omega_{\ell} | A_{\mu}^{a}| \Omega \rangle \, 
\frac{\delta_{ -{\vec q},{\vec p}_{\ell} }\delta( \varepsilon_{\ell}-\omega)}{2\varepsilon_{\ell}} ,
\label{eq:rhom}
\end{eqnarray}
where $|\Omega_{\ell}\rangle$ are the
eigenstates of the QCD Hamiltonian  normalized as 
$\langle \Omega_{\ell} | \Omega_{\ell^{\prime}} \rangle = 2 \varepsilon_{\ell} (2\pi)^{3}$ $V \delta_{\vec p_{\ell}, \vec p_{\ell^{\prime}}}$
with the eigenvalue $\varepsilon_{\ell}$ measured from the ground state and the spatial volume $V$.
In the infinite volume limit, 
the summation should be replaced as
$\sum_{\ell} \rightarrow \int d\varepsilon\, n(\varepsilon)$ with 
the density of states $n(\varepsilon)$.

Since the states $|\Omega \rangle$ and $|\Omega_l \rangle $ have
isospin 0 and 1, respectively, we put the isospin label $a\, (=1,2,3)$ 
explicitly to the latter state as 
$| \Omega_{\ell}^{a}\rangle$ for later convenience.
In the soft limit $q_{\mu} \rightarrow 0$, 
we have zero modes denoted by $\ell=\alpha$ with the 
property  $\varepsilon_{\alpha} \rightarrow 0$ as $\vec q \rightarrow 0$,
and the non-zero modes denoted by $\ell=\beta$ with the
property  $\varepsilon_{\beta} \neq 0$ at $\vec q = 0$.\footnote{Strictly
speaking, we should introduce a small quark mass $m_q$ and take the
limit $m_q \rightarrow 0$ after  $V \rightarrow \infty$ to make 
the whole procedure well-defined.}   
Leaving the discussion on the physical content of these states to the next
section, we focus on the symmetry aspect in the nuclear medium.

With the vector $n_{\mu}$ characterizing Lorentz frame of the nuclear matter,  
the matrix elements of $\phi_{5}^{a}$ and  $A_{\mu}^{a}$ for
the states $|\Omega\rangle$ and $|\Omega_{\alpha}^{a}\rangle$ are generically
written as 
\begin{eqnarray}
\langle \Omega^{b}_{\ell}(k) | \phi^{a}_{5}(x) | \Omega \rangle &=& \delta^{ab} 
G_{\ell}^{*1/2}
e^{ik\cdot x}, \label{eq:medZ} \\
\langle \Omega | A_{\mu}^{a}(x) | \Omega^{b}_{\ell}(k) \rangle &=& 
i \delta^{ab} [ n_{\mu} (n\cdot k) N_{\ell}^* +  k_{\mu} F^{*}_{\ell}] e^{-ik\cdot x}.
\label{eq:matAmu} 
\end{eqnarray}
Here the space-time independent constants,
$G^{*}_{\ell}$ (pseudo-scalar coupling constant),
and  $N_{\ell}^*$ and $F^{*}_{\ell}$ (axial-vector coupling
constants) are 
the functions of $k\cdot n$ and $k^{2}$.
The factor $n\cdot k$ in front of $N_{\ell}^*$ is introduced to
make  $N_{\ell}^*$ and $F^{*}_{\ell}$ have the same dimension.
Because of the conservation law of the axial current $\partial \cdot A =0$ in the 
chiral limit,  $N_{\ell}^*$ and $F^{*}_{\ell}$ are not independent: 
$(k\cdot n)^{2} N_{\ell}^* + k^{2} F^{*}_{\ell} = 0$. 
Taking the rest frame of the nuclear matter where $n_{\mu}=(1,0,0,0)$,
we have
\begin{equation}
\varepsilon_{\ell}^{2} (N_{\ell}^*+F^{*}_{\ell}) - {\vec k}^{2} F^{*}_{\ell} = 0 ,
\label{eq:CACforNF}
\end{equation}
with $k=(\varepsilon_{\ell}, {\vec k})$.
According to this relation, the axial-vector coupling of the 
non-zero modes $N_{\beta}^{*}+F_{\beta}^{*}$ 
should vanish in the limit $\vec k \rightarrow 0$. 

The correlator~\eqref{eq:disp}
can be calculated with the matrix elements \eqref{eq:medZ} and \eqref{eq:matAmu}:
For example, 
the spectral function $\sigma_{\mu}^{+}$
defined in \eqref{eq:rhom}
is evaluated to be 
\begin{eqnarray}
\ \ \ \ \ \  \  & & i \delta^{ab}
\sum_{\ell} \frac{\omega \varepsilon_{\ell} N_{\ell}^*
+ (\omega \varepsilon_{\ell} - {\vec q}^{2}) 
F_{\ell}^{*}}{2 \varepsilon_{\ell}
(\omega - \varepsilon_{\ell} + i \epsilon )} G_{\ell}^{*1/2}  \nonumber \\
& & \ \ \ \ \ \ \ \ \ \ \ \ \ \ \ \ 
\xrightarrow[q_{\mu} \rightarrow 0]{ } \ \ \ 
 \frac{i}{2} \delta^{ab} \sum_{\alpha}(N_{\alpha}^* +F_{\alpha}^{*}) G^{*1/2}_{\alpha} , 
\end{eqnarray}
where we have used Eq.\eqref{eq:CACforNF} together with the 
fact that $N_{\ell}^{*}+F_{\ell}^{*}$
can have finite values only for the zero-modes ($\ell=\alpha$).

 Note that the result
is independent of how to take the soft limit, $q_{\mu} \rightarrow 0$. 
The contribution from $\sigma^{-}_{\mu}$ 
is found to be the complex conjugate of
$\sigma^{+}_{\mu}$. 
Thus the hadronic contribution to the correlator \eqref{eq:Aphicorr}
in the soft limit 
becomes
\begin{equation}
\Pi_5^{ab}(0) 
=  i\delta^{ab} \sum_{\alpha} {\rm Re} \left[ (N_{\alpha}^* + F_{\alpha}^{*}  ) 
G^{*1/2}_{\alpha} \right] . \label{eq:FD1st}
\end{equation}
where the summation is taken for all the zero modes. 

Equating Eqs.\eqref{eq:FD2nd} and \eqref{eq:FD1st}, 
we finally obtain an exact sum rule in the chiral limit
which is valid for {\em all densities},
\begin{equation}
\sum_{\alpha} {\rm Re} \left[ (N_{\alpha}^* + F_{\alpha}^{*}  )   G^{*1/2}_{\alpha} \right]
= -  \langle\bar q q  \rangle^{*}  \ .
\label{eq:mFZqq}
\end{equation}
At zero density, only the pion 
contributes to the sum rule and  Eq.~\eqref{eq:mFZqq}
reduces to the well-known Glashow-Weinberg relation,
$F_{\pi} G_{\pi}^{1/2}=-\langle\bar qq \rangle$ \cite{Glashow:1967rx}. 
On the other hand, at finite density,
the left hand side of  Eq.~\eqref{eq:mFZqq} may receive various contributions
not only from the in-medium pion but also from other collective excitations
in  
nuclear matter, 
so that full dynamical treatment of the many body system would be
necessary. However, the 
discussion becomes  much simpler when 
the linear density approximation is adopted
as will be shown in the next section. 

Before closing this section, let us 
consider a zero mode 
with a linear dispersion relation  $\varepsilon_{\alpha} = v_{\alpha} |\vec k| $
for small momenta.  In this case, the matrix element \eqref{eq:matAmu}
can be written in a simpler form. Let us 
introduce the temporal and spatial ``decay constants" as 
\begin{eqnarray}
\ \ \ \ \ \ \ \ \ \ \ \ \ \ \ 
& & \langle \Omega | A_{0}^{a}(0) | \Omega_{\alpha}^{b}(k) \rangle =   i \delta^{ab}
\varepsilon_{\alpha} \,  F^{t}_{\alpha}  \ ,
 \label{eq:medFt} \\
\ \ \ \ \ \ \ \ \ \ \ \ \ \ \ 
& & \langle \Omega | A_{i}^{a}(0) | \Omega_{\alpha}^{b}(k) \rangle =    i \delta^{ab}
 k_{i}\,  F^{s}_{\alpha} \   . \label{eq:medFs}
\end{eqnarray}
Then, by using the dispersion relation together with 
Eq.~\eqref{eq:CACforNF} and $n_{\mu}=(1,0,0,0)$,
we have 
\begin{equation}
F_{\alpha}^{s} = F^{*}_{\alpha}, \ \ \
F^t_{\alpha}   = N_{\alpha}^* + F_{\alpha}^* =  F^{s}_{\alpha}/v_{\alpha}^{2}.
\end{equation}  
Here it should be noticed that 
the relation $F^t_{\pi}=  F^{s}_{\pi}/v_{\pi}^{2}$ 
for the pionic mode  {\em at finite temperature} 
is  derived in 
\cite{Pisarski:1996mt,Harada:2002zh}.

\section{In-medium Glashow-Weinberg relation at low density}
\label{sec:GW}

\begin{figure}
\epsfxsize=8.5cm
\begin{center}
\epsfbox{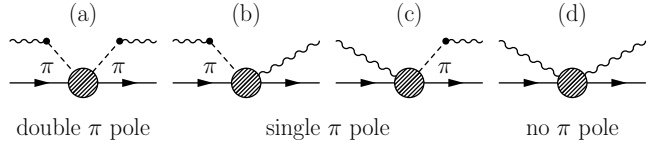}
\caption{
Diagrams for the $\pi$N scattering amplitude induced by the external 
fields, $A_{\mu}^{a}$ and $\phi_{5}^{b}$. The solid, dotted and wavy lines
denote nucleon, pion and external fields, respectively. 
} \label{fig1}
\end{center}
\end{figure}

The exact sum rule in the chiral limit, Eq. (\ref{eq:mFZqq}),
is reduced to a simple form at low density.
The expectation value of an arbitrary operator ${\cal O}$
in nuclear matter up to the linear term in density reads
\begin{equation}
\langle \Omega | {\cal O} | \Omega \rangle \simeq
\langle 0 | {\cal O} | 0 \rangle + \rho \langle N | {\cal O} | N \rangle .
\label{eq:linear-density}
\end{equation}
For $\Pi_5^{ab}(q)$ in Eq.(\ref{eq:Aphicorr}),
the first term in the right hand side of Eq.(\ref{eq:linear-density})
is dominated by the massless pion in the chiral limit.
The second term corresponds to
the forward scattering amplitude of the external fields ($A_{\mu}^a$ and
$\phi_5^b$) with the nucleon multiplied by $q^{\mu}$.

As shown in Fig.\ref{fig1}, there are four different contributions
to the forward amplitude; (a) the term with double $\pi$ pole where
both $A_{\mu}^a$ and $\phi_5^b$ couple directly to the pion,
(b,c) the term with single $\pi$ pole where either $A_{\mu}^a$ or
$\phi_5^b$
couples directly to the pion, and (d) the term without $\pi$ pole
where the external fields couple directly to the nucleon.

It is easy to see that the $p$-wave coupling of the pion or the external 
fields
to the nucleon gives vanishing contribution to the forward amplitude in
(a-d).
On the other hand, the $s$-wave coupling in (a,b,c) can leave finite
contributions
even in the soft limit due to the pion pole(s).
In particular, particle-hole excitations do not contribute to the 
sum rule in the linear-density approximation.
This leads to the conclusion that
only the pionic mode with possible medium modification
contributes to the sum rule in the previous section at low density:
\begin{equation}
F^t _{\pi} G^{*1/2}_{\pi} = -\langle \bar q q \rangle^{*} .
\label{eq:piFZqq}
\end{equation}
Note that Eq.(\ref{eq:piFZqq}) is only valid if both sides are expanded
up to $O(\rho)$.

Taking the ratio of Eq.\eqref{eq:piFZqq} and its counterpart at zero density,
we find the scaling law:
\begin{equation}
\left( \frac{ F^t _{\pi}}{ F_{\pi}} \right)
Z_{\pi}^{*1/2}  = 
\frac{\ \langle {\bar q}q \rangle^{*}}{\langle\bar qq \rangle}\ ,
\label{eq:newrelat}
\end{equation} 
{with the in-medium wave function renormalization 
$Z_{\pi}^{*} \equiv G_{\pi}^{*}/G_{\pi}$.}
As will be shown in the next section, in-medium change of 
{$Z_{\pi}^{*}$ from 1} 
can be evaluated from the isosinglet pion-nucleon
scattering amplitude, while  
$F_{\pi}^t/F_{\pi}$ is related to the pion-nucleus isovector scattering
lengths through the in-medium TW relation.  
Therefore, Eq.(\ref{eq:newrelat}) gives a direct link
between the in-medium modification of the quark condensate and that
of the pion decay constant\footnote{There is  also a well-known
 low-energy theorem at low density which relates the 
 in-medium modification of the quark condensate to the pion-nucleon
  $\sigma$-term \cite{Drukarev:1991fs};
$\langle {\bar q}q \rangle^{*}/{\langle\bar qq \rangle}= 
1 - \sigma_{\pi N}\rho /(F_{\pi}^2 m_{\pi}^2) $ with
$\sigma_{\pi N} \simeq 45$ MeV. \label{footnote2}}.

Alternative relation between  $\langle {\bar q}q \rangle^{*}$ and $F_{\pi}^t$
is obtained  by taking the matrix element of the PCAC relation\\ 
$\partial\cdot A^{a}(x) = 2 m_{q} \phi_{5}^{a}(x)$ slightly away from the 
chiral limit. With the matrix elments given in
Eqs.~(\ref{eq:medZ}) and  (\ref{eq:medFt}) with $\alpha=\pi$, 
we have $(\varepsilon^{2}-v_{\pi}^{2} {\vec k}^{2}) F^{t}_{\pi} = 2m_{q} G^{* 1/2}_{\pi}$,
where $ F^{t}_{\pi}$ and $G^{*}_{\pi}$ are the values in the chiral and soft limit.
Since the in-medium pion mass is given by
$\varepsilon^{2} = \pimassrhosq +  v_{\pi}^{2} {\vec k}^{ 2} + O({\vec k}^{ 4}) $, we have
$\pimassrhosq F_{\pi }^{t}  = 2 m_{q} G^{*1/2}_{\pi}$.  
Combining this with Eq.(\ref{eq:mFZqq}), we obtain 
\begin{eqnarray}
\label{eq:GOR}
\ \ \ \ \ \ \ \ \ \ \ \ \ \ \ \ \ \ \ \ 
& &  
( F^{t}_{\pi})^{2} \pimassrhosq   = - 2 m_{q} \langle \bar q q \rangle^{*} ,\\
\label{eq:GOR-ratio}
\ \ \ \ \ \ \ \ \ \ \ \ \ \ \ \ \ \ \ \ 
& & 
\left( \frac{ F^t _{\pi}}{ F_{\pi}} \right)^2
\left( \frac{m^{*}_{\pi}}{m_{\pi}} \right)^{2} 
  = 
\frac{\ \langle {\bar q}q \rangle^{*}}{\langle\bar qq \rangle} .
\end{eqnarray}
Eq.(\ref{eq:GOR}) is the in-medium generalization of the 
Gell-Mann--Oakes--Renner relation \cite{GellMann:1968rz}
and was derived before
in the Nambu--Jona-Lasinio model~\cite{Hatsuda:1985ey} and
in chiral perturbation theory~\cite{Thorsson:1995rj,Pisarski:1996mt}. 
Theoretically,  Eq.(\ref{eq:newrelat}) is
equivalent to Eq.(\ref{eq:GOR-ratio}). Experimentally,
the information on $m_{\pi}^*$ is necessary to check the latter sum rule,
which is
relatively difficult.  

\section{In-medium wave-function renormalization}
\label{sec:WF}

Let us consider the relation between the pion wave function renormalization
constant 
{$Z_{\pi}^*$}
and the isospin singlet pion-nucleon scattering amplitude.
For this purpose, we introduce the off-shell $\pi$N amplitude near the
pion pole through the operator $\phi_5^a(x)$, as is done in
\cite{Weinberg:1966kf},
\begin{eqnarray}
& & {\cal T}_{\pi N}^{ab}(\nu,\nu_B;m_{\pi}) = \delta^{ab}\ {\cal T}^{(+)}
+ \frac{1}{2} [\tau^a,\tau^b] \ {\cal T}^{(-)} 
\nonumber \\ 
& & \equiv  \frac{i}{G_{\pi}} q^{2}q^{\prime2}
\int d^{4}x  \, 
e^{iq\cdot x} \langle N(P') | {\rm T} [\phi_{5}^{a}(x) \phi_{5}^{b}(0)] | N(P) \rangle 
\label{eq:defpiNamp}
\end{eqnarray}
with the in-coming (out-going) pion momentum $q$ ($q^{\prime}$)
and the kinematical variables defined as
$\nu \equiv P \cdot (q+q^{\prime})/ (2 M_{N})$ and 
$ \nu_B  \equiv  - q \cdot q^{\prime} /(2M_N)$. 
In the forward limit $q^{\prime} \rightarrow q$ with $\vec{q}=0$,
the scattering amplitude is a function solely of $\omega$.
Thus, the isospin singlet amplitude for small $\omega$ is
written as 
\begin{equation}
{\cal T}^{(+)}(\omega;m_{\pi}) \simeq \alpha + \beta \omega^{2}
\label{eq:piNamp}
\end{equation}
Thanks to the special off-shell extrapolation given in Eq.(\ref{eq:defpiNamp}),
the coefficients $\alpha$ and $\beta$ can be evaluated as follows \cite{Weinberg:1966kf}:
(i) At the off-shell Weinberg point, we have ${\cal T}^{(+)}(0;m_{\pi}) = \alpha =
-\sigma_{\pi N}/F_{\pi}^{2}$ with the $\pi N$ sigma term 
$\sigma_{\pi N}\simeq 45 {\rm MeV}$~\cite{Sainio:2001bq}. 
(ii)  At the on-shell threshold,  we have
${\cal T}^{(+)}(\omega=\pimass;m_{\pi})
= 4\pi (1+\pimass/m_{N}) a_{\pi N}$ with the scattering length
$a_{\pi N}=(0.0016\pm 0.0013)m_{\pi}^{-1} $~\cite{Schroder:1999uq}.
Combining these,  we obtain 
\begin{equation}
\beta \simeq  \frac{\sigma_{\pi N}}{F_{\pi}^{2}\pimass^{2}}
+ \left( 1+\frac{m_{\pi}}{m_N} \right) \frac{4\pi a_{\pi N}}{m_{\pi}^2} .
\label{eq:betabeta}
\end{equation}
Numerically, the first term in the right hand side of Eq.(\ref{eq:betabeta})
dominates over the  second term and we find
$\beta  = 2.17 \pm 0.04 \ {\rm fm}^3$.

To obtain a relation between $\beta$ and 
{$Z_{\pi}^*$,} 
we now consider the correlation of $\phi_5^a$ 
in  symmetric nuclear matter in the chiral limit
expanded up to linear in density according to Eqs.(\ref{eq:linear-density}),
(\ref{eq:defpiNamp}) and (\ref{eq:piNamp});
\begin{eqnarray}
D^{ab}(q) &=& 
\int d^{4} x\, e^{i q \cdot x} \langle \Omega | {\rm T}[ \phi_{5}^{a}(x) \phi_{5}^{b}(0)] | \Omega \rangle\  
\label{eq:medPcor} \\ 
\! \! \!  \! \! & \xrightarrow[{\bf q}=0]{ } & \ \ i \delta^{ab} G_{\pi} \left[
\frac{1}{\omega^{2}} - \frac{1}{\omega^{2}}   {\cal T}^{(+)}(\omega;0) \rho 
 \frac{1}{\omega^{2}} 
\right] ,
\label{eq:medprop} \\
\! \! \!  \! \! & = &  \ \ i \delta^{ab} \frac{G_{\pi} (1-\beta \rho)}{\omega^{2}}
= i \delta^{ab} \frac{G_{\pi}^*}{\omega^{2}}  .
\end{eqnarray}
Then,  we obtain
\begin{eqnarray}
\ \ \ \ \ \ 
Z_{\pi}^{*1/2} \equiv \left(  \frac{G^{*}_{\pi}}{G_{\pi}} \right)^{1/2}
=  1 - \gamma \frac{\rho}{\rho_0},
\label{eq:wfren} 
\end{eqnarray}
with $\gamma = \beta \rho_0/2 \simeq 0.184$.
Notice that the reduction of 
{$Z_{\pi}^*$} 
in the nuclear medium 
given in Eq.\eqref{eq:wfren}  stems  solely 
from the $s$-wave pion-nucleon interaction.

\section{In-medium Tomozawa-Weinberg relation}
\label{sec:TW}

In-medium pion 
properties are conventionally expressed in terms of  the pion-nucleus optical potential.
For instance, the $s$-wave potential for $\pi^{-}$ is parametrized as 
\cite{Kolomeitsev:2002gc}
\begin{eqnarray}
\label{eq:b0-b1}
& & 2\pimass  U_{\rm s} = - 4 \pi \left[ 1 + {\textstyle \frac{\pimass}{m_{N}}}\right] 
\ \left( b_{0}^*(\rho) \rho - b_{1}^*(\rho) \delta\rho \right) , \\
\label{eq:Tpm}
&  & \ \ \ =  - {\cal T}^{(+)*}(\omega=m_{\pi};m_{\pi}) \rho
 - {\cal T}^{(-)*}(\omega=m_{\pi};m_{\pi})
\delta \rho ,
\end{eqnarray}
with the isoscalar density $\rho = \rho_{p} + \rho_{n}$ and the 
isovector density $\delta \rho = \rho_{p}-\rho_{n}$.
The parameters $b_{0}^*$ and $b_{1}^*$ represent
the pion-nucleus scattering lengths in isoscalar and isovector channels, 
respectively. ${\cal T}^{(\pm)*}(\omega;m_{\pi})$ are the isoscalar and isovector
pion-nucleus scattering amplitude at zero spatial momentum.
The value of $b_{1}^*$ extracted from the pionic atom and $\pi^{-}$-nucleus scattering data 
is larger than that in the vacuum, which indicates an enhanced repulsion 
from nuclei  \cite{Suzuki:2002ae,Friedman:2004jh}. 
This extra repulsion expressed by $b_{1}^*$ is 
interpreted as originating from the 
in-medium reduction of $F_{\pi}^t$ through
the in-medium generalization of the 
Tomozawa-Weinberg relation \cite{Kolomeitsev:2002gc,Weise:2005ss}.

Let us derive a relation between $b_1^*$ and $F_{\pi}^t$ 
in a way parallel to the derivation of 
{$Z_{\pi}^*$} 
in the previous section using 
current commutation relations.
We consider the following axial vector correlator
in a slightly asymmetric nuclear matter $| \Omega^{\prime} \rangle $ in the chiral limit:
\begin{equation}
\Pi_{\nu}^{ab}(q) =   \int d^{4}x \ e^{i q\cdot x} \partial^{\mu} 
\langle \Omega^{\prime} | {\rm T} [A_{\mu}^{a}(x)  A_{\nu}^{b}(0)] | \Omega^{\prime} \rangle \ . 
\label{eq:AAcorr}
\end{equation}
Then using the current conservation $\partial \cdot A=0 $ and
the relation 
$ \int d^4 x\  \partial^\mu {\rm T} [A_\mu^a(x) A_{\nu}^b(0)] 
= [Q_5^a, A_{\nu}^b(0)] = i \epsilon^{abc} V^{c}_{\nu}(0)$ 
satisfied in the chiral limit,
we obtain the sum rule  
\begin{equation}
\Pi_{0}^{ab}(0) 
= i \epsilon^{ab3} \langle \Omega'|V_0^3 | \Omega' \rangle 
\simeq i \epsilon^{ab3} \frac{1}{2} \delta \rho .
\label{eq:MTVC}
\end{equation}
On the other hand, Eq.\eqref{eq:AAcorr} at the chiral and soft limits  is saturated
with linear in isovector density as   
\begin{equation}  
\Pi_0^{12}(q) \xrightarrow[{\bf q}=0]{ }
i \omega \left[ \frac{ \omega F_{\pi}^t}{\omega^2}
\cdot {\cal T}^{(-)*}(\omega;0) \delta \rho \cdot
\frac{ \omega F_{\pi}^t }{\omega^2}
\right] ,
\label{eq:IVsoftlimit}
\end{equation}  
Comparing Eqs.~\eqref{eq:MTVC} and \eqref{eq:IVsoftlimit}, we find
\begin{equation}
\label{eq:mTW}
{\cal T}^{(-)*}(\omega;0) 
\simeq  \frac{\omega}{2(F_{\pi}^{t})^{2}}  .
\end{equation}
This is an in-medium generalization of the Tomozawa-Weinberg relation
\cite{Tomozawa:1966jm,Weinberg:1966kf} and was
obtained before in Ref.\cite{Kolomeitsev:2002gc} 
with chiral perturbation approach.

Using the definitions Eqs.(\ref{eq:b0-b1},\ref{eq:Tpm})
together with  Eq.(\ref{eq:mTW}),  
we obtain a formula relating the in-medium change of the 
isovector scattering length and the in-medium
change of the pion decay constant in the chiral limit:
\begin{equation}
\frac{b_{1}}{b_{1}^*}= \left( \frac{F^{t}_{\pi}}{F_{\pi}} \right)^2 .
\label{eq:TW}
\end{equation}

\section{In-medium quark condensate}
\label{sec:QC}

Now,  inserting Eqs.\eqref{eq:wfren} and \eqref{eq:TW} into Eq.\eqref{eq:piFZqq},
we arrive at one of the central results in this work,  
\begin{equation}
\frac{\ \langle  \bar q q \rangle^{*}}{\langle\bar qq \rangle} \simeq
\left( \frac{b_1}{b_1^*} \right)^{1/2}
\left( 1- \gamma \frac{\rho}{\rho_0} \right),
\label{eq:final}
\end{equation}
which directly relates the in-medium 
quark condensate with  the observables related to the pion in 
nuclei.  The deeply bound pionic atom data suggest the repulsive enhancement 
of $b_{1}^{*}$~\cite{Suzuki:2002ae}. The $\pi$N scattering data tell 
that $\gamma >0$.
Thus, Eq.\eqref{eq:final} implies that these experimental
facts give  a direct evidence of the reduction of the quark condensate
in nuclear medium.
Quantitatively, the 
experimental value of ${b_1}/{b_1^*}$ is obtained as
$0.79\pm 0.05$ at the effective density $\rho \approx 0.6
\rho_{0}$ in deeply bound pionic atoms~\cite{Suzuki:2002ae}.
With this value and $\gamma=0.184$ estimated in Sec.~\ref{sec:WF}
together with the linear density approximation, we
find for the  ratio of the quark condensates
$\langle \bar q q \rangle^{*} / \langle\bar qq \rangle
\simeq  1 - 0.37 \rho/\rho_{0}$. We also evaluate this ratio with 
{${b_1}/{b_1^*}=0.75$}
obtained in elastic $\pi$-nucleus scatterings 
\cite{Friedman:2004jh} 
assuming the effective density $\rho \approx \rho_{0}$. 
The result is
{$\langle \bar q q \rangle^{*} / \langle\bar qq \rangle
\simeq 1- 0.43 \rho/\rho_{0}$.}
These numbers are consistent with 
that given by the formula expressed with the 
pion-nucleon $\sigma$-term (see footnote \ref{footnote2}.)
{
For further quantitative argument, one has to take into account
explicit chiral symmetry breaking effects and higher density contributions.
}

Here we emphasize 
that the renormalization of the pion field is inevitable
in describing the pion dynamics when
partial restoration of chiral symmetry occurs.
This is understood 
clearly 
in the 
context of the chiral effective theory.
The chiral effective theory is based on a consistent decomposition 
of the field variables on the chiral manifold with the original symmetry to
radial ($\sigma$)  and angular (pionic, i.e., Nambu-Goldstone) ones.
For dynamics in the vacuum, the relevant degree of freedom 
is the massless angular mode expressed by the dimensionless chiral field $U$.
Since the pion field has the dimension of energy and  
the order parameter of the dynamical symmetry breaking 
provides the only relevant energy scale 
in the chiral limit, the pion field should be normalized by the quark condensate. 
Therefore, when partial restoration of the chiral symmetry 
takes place, 
the pion field is necessarily renormalized according
to the reduction of the quark condensate~\cite{Jido:2000bw};
see also Sec.\ref{sec:ChPT}.

\section{The sum rule beyond the chiral limit}
\label{sec:offCL}
Let us generalize the in-medium sum rule Eq.(\ref{eq:mFZqq})
in the chiral limit to the case with  a finite quark mass $m_q$.
From the PCAC  relation $\partial \cdot A^{a}(x) = 2 m_{q} \phi_{5}^{a}(x)$,
Eq.(\ref{eq:FD2nd}) is easily generalized to
\begin{equation}
\Pi_{5}^{ab}(0)-2m_{q} D^{ab}(0) = -i \delta^{ab} \langle \bar q q \rangle^{*}
\label{eq:preSR}
\end{equation}
where $D^{ab}(q)$ is the correlation function of the pseudoscalar density 
$\phi_{5}^{a}$ given in Eq.(\ref{eq:medPcor}).

Also, the definition of the coupling constants,
Eqs.(\ref{eq:medZ}) and (\ref{eq:matAmu}), together with the PCAC relation
lead to the generalization of Eq.(\ref{eq:CACforNF}),
\begin{equation}
\varepsilon_{\ell}^{2} (N_{\ell}^*+F^{*}_{\ell}) 
- {\vec k}^{2} F^{*}_{\ell} = 2m_{q} G_{\ell}^{*1/2} .
\label{eq:PCACforNF}
\end{equation}

The hadronic matrix elements in the left hand side of Eq.(\ref{eq:preSR})
can be evaluated as before and we find the sum rule away from the 
chiral limit, 
\begin{equation}
 \sum_{\ell} {\rm Re}\left[ (N_{\ell}^* +F_{\ell}^{*})\, G^{*1/2}_{\ell} \right]
 = - \langle \bar q q \rangle^{*}, \label{eq:SRoffCL}
\end{equation}
where the summation is taken over all states coupled to 
$A_{\mu}$ and $\phi_{5}$, and  the matrix elements are evaluated 
at $k_{\ell}=(\varepsilon_{\ell}(\vec k =0), \vec 0)$. 
Modes $\ell$ may be classified into would-be zero modes ($\ell=\alpha$)
and would-be non-zero modes ($\ell =\beta$). 
Then, eq.(\ref{eq:PCACforNF}) shows that, in the soft limit,
$N_{\alpha}^* +F_{\alpha}^{*} = {\rm const.} + O(m_q)$ for
 would-be zero-modes, while 
 $N_{\beta}^* +F_{\beta}^{*} = O(m_q)$ for would-be non-zero modes.
Namely, the quark mass correction to 
$\langle \bar q q \rangle^{*} $ in the rhs of Eq.(\ref{eq:SRoffCL})
receives  both effects. To evaluate them, 
detailed hadronic model calculations such as given in  
Ref.~\cite{Kaiser:2007nv} are needed.

\section{Relation to the chiral effective theory}
\label{sec:ChPT}

So far, we have derived the in-medium sum rule Eq.(\ref{eq:mFZqq})
based only on general operator relations in QCD 
without assuming any hadronic descriptions.
The sum rule can be used for checking consistency 
of theoretical models with the fundamental symmetry of QCD,
and also for experimental confirmations of partial restoration of chiral 
symmetry in nuclear medium, once the matrix elements of the currents
are experimentally extracted, as discussed in section \ref{sec:QC}.

Let us demonstrate here how the in-medium GW relation Eq.(\ref{eq:piFZqq}) 
is expressed in terms of the low energy constants in chiral perturbation theory. 
An effective Lagrangian for the pion in nuclear medium is obtained in the mean field 
approximation of the nucleon field \cite{Thorsson:1995rj}:
\begin{eqnarray}
 {\cal L}^{\rm eff} &=& \left(\frac{f^{2}}{4} + \frac{c_{3}}{2}\rho \right) 
 {\rm Tr}[D_{\mu} U D^{\mu} U^{\dagger}] 
 +  \frac{\tilde c_{2}}{2} \rho \,
 {\rm Tr}[D_{0} U D_{0} U^{\dagger}] 
 \nonumber \\ &&
 +  \left(\frac{f^{2}}{4} + c_{1}\rho \right) 
  {\rm Tr}[U^{\dagger} \chi + \chi^{\dagger}U]  \label{eq:effLag},
\end{eqnarray}
where $\rho=\rho_{p} + \rho_{n}$ is the isoscalar density , 
$c_{i}$ $(i=1,2,3)$ are the low energy constants with
$\tilde c_{2} = c_{2} -g_{A}^{2}/(8\Nmass)$ and $f$ is the tree-level 
pion decay constant in the vacuum in the chiral limit. 
We also introduce the chiral vector currents, $\ell_{\mu}$ and $r_{\mu}$, in the covariant derivative
$
 D_{\mu} U = \partial_{\mu} U  - i r_{\mu} U  + i  U \ell_{\mu},
$
and the scalar and pseudoscalar external fields, $s$ and $p$, in
$
 \chi = 2B_{0}(s + i p) 
$
with a normalization constant $B_{0}$.
The chiral field $U$ is given by 
\begin{equation}
U=\exp[i \frac{\pi^a \cdot \tau^a}{f^*}].
\end{equation} 
The $f^*$ represents the normalization of the pion field $\pi^a$
and is left as a parameter to be fixed so as to normalize the pion 
kinetic term properly~\cite{Jido:2000bw}. Expanding the effective Lagrangian~(\ref{eq:effLag}) in terms of the pion field,  we find the normalization in the linear-density 
approximation to be
\begin{equation}
f^* =  f \left[ 1 + \frac{\rho}{f^{2}} 
\left(c_{3} +  \tilde c_{2} \right)\right]  . \label{eq:ChPTnorm}
\end{equation}

The axial current and the pseudoscalar density 
in the effective Lagrangian are evaluated
through the formulas
$ A_{\mu}^{a}(x)  = - (
 {\partial {\cal L}^{\rm eff}}/{\partial r^{\mu}_{a}} - 
 {\partial {\cal L}^{\rm eff}}/{\partial \ell^{\mu}_{a}})/2$
and
$   \phi^{a}(x) =   {\partial {\cal L}^{\rm eff}}/{\partial p^{a}}$.
Evaluating the matrix elements of the axial current and the pseudoscalar 
density for the in-medium one-pion state in the tree approximation, we obtain
\begin{equation}
F_{\pi}^{t} = \frac{f^{2} + (2 c_{3}+2 \tilde c_{2}) \rho}{f^*},
\label{eq:ChPTft}
\ \ \
G_{\pi}^{* 1/2} =   2B_{0} \frac{f^{2} + 4c_{1}\rho}{f^*} .
\end{equation}
Thus, combining Eq.(\ref{eq:ChPTft}) 
together with
the pion field normalization (\ref{eq:ChPTnorm}),
we obtain
\begin{equation}
F_{\pi}^{t} G_{\pi}^{* 1/2} =  -2B_{0} (f^{2} + 4c_{1}\rho),
\end{equation}
in the linear density approximation. 
This is exactly the same formula as Eq.\eqref{eq:piFZqq}
for the medium modification of the 
quark condensate,
since $\langle \bar q q \rangle^{*} =
\langle - \partial {\cal L}^{\rm eff}/{\partial s^{0}}
\rangle^{*} =  -2B_{0} (f^{2} + 4c_{1}\rho)$.

To evaluate the sum rule beyond the linear density approximation,
one has to calculate the matrix elements with taking into
account of dynamical properties of the pion and nucleons in nuclear medium.
For this purpose, the chiral effective theory is one of the powerful
tool which  gives systematic description in terms of density and chiral 
symmetry breaking off soft limit.

\section{Summary}
\label{sec:summary}

In this paper, we have derived various in-medium sum rules 
in a model-independent way by  exploiting  operator relations and 
chiral symmetry in QCD.
We have found an exact sum rule Eq.\eqref{eq:mFZqq} valid for any density in the 
chiral limit.  Also, a new scaling law Eq.\eqref{eq:newrelat}
as a generalization of the Glashow-Weinberg relation is deduced 
from the exact sum rule at low density, in which the in-medium
quark condensate,  the in-medium pion decay constant 
and the in-medium wave-function renormalization of the pion 
field  are  related. 
  With the information of the 
$\pi$N scattering data and the in-medium Tomozawa-Weinberg relation Eq.(\ref{eq:TW}),
 the quark condensate is related directly to the 
isovector $\pi$-nucleus scattering length as shown in Eq.(\ref{eq:final}).

Our formulas should be useful to check the consistency among in-medium
pion properties at low density from the point of view of chiral symmetry,
and to extract the in-medium quark condensate.
Generalization of our approach to the pion-pion interaction inside
 a nuclear medium following the idea of \cite{Jido:2000bw} is one 
of the interesting future problems to be examined and  may  
provide us with a further insight into the in-medium pions and
possible partial restoration
of chiral symmetry in nuclei.   
It is also an interesting and important task
to 
estimate contributions beyond 
the linear density approximation 
~\cite{Kaiser:2007nv,Meissner:2001gz,Doring:2007qi}.

\section*{Acknowledgments}
D.J.\ thanks W.\ Weise and S.\ Hirenzaki for their continuous encouragements 
and helpful discussions. 
This work was partially supported  by 
a Grant-in-Aid
for Scientific Research by Monbu-Kagakusyo of Japan
(Nos.17540250, 18042001, 18540253, 20028004, 20540265, 20540273) and was
done under  Yukawa International Program for Quark-Hadron Sciences. 

\end{document}